\documentclass[runningheads]{llncs}
\usepackage[T1]{fontenc}
\usepackage{orcidlink}
\usepackage{graphicx}
\begin{document}
\title{LLM-Driven Augmented Reality Puppeteer: Controller-Free Voice-Commanded Robot Teleoperation}
%
%
\author{
Yuchong Zhang\inst{1}\orcidID{0000-0003-1804-6296}\thanks{Contributed equally to this work.} \and
Bastian Orthmann\inst{1}\orcidID{0000-0001-8542-255X}\thanks{Contributed equally to this work.} \and
Michael C. Welle\inst{1}\orcidID{0000-0003-3827-3824} \and Jonne Van Haastregt \inst{1} \and
Danica Kragic \inst{1}\orcidID{0000-0003-2965-2953}}
\authorrunning{Zhang and Orthmann et al.}
%
\institute{KTH Royal Institute of Technology Stockholm, Sweden\\
\email{\{yuchongz,orthmann,mwelle,dani\}@kth.se; \\ jonnevanhaastregt@gmail.com}}
\maketitle              

\begin{figure}
\centering
\includegraphics[width=\textwidth]{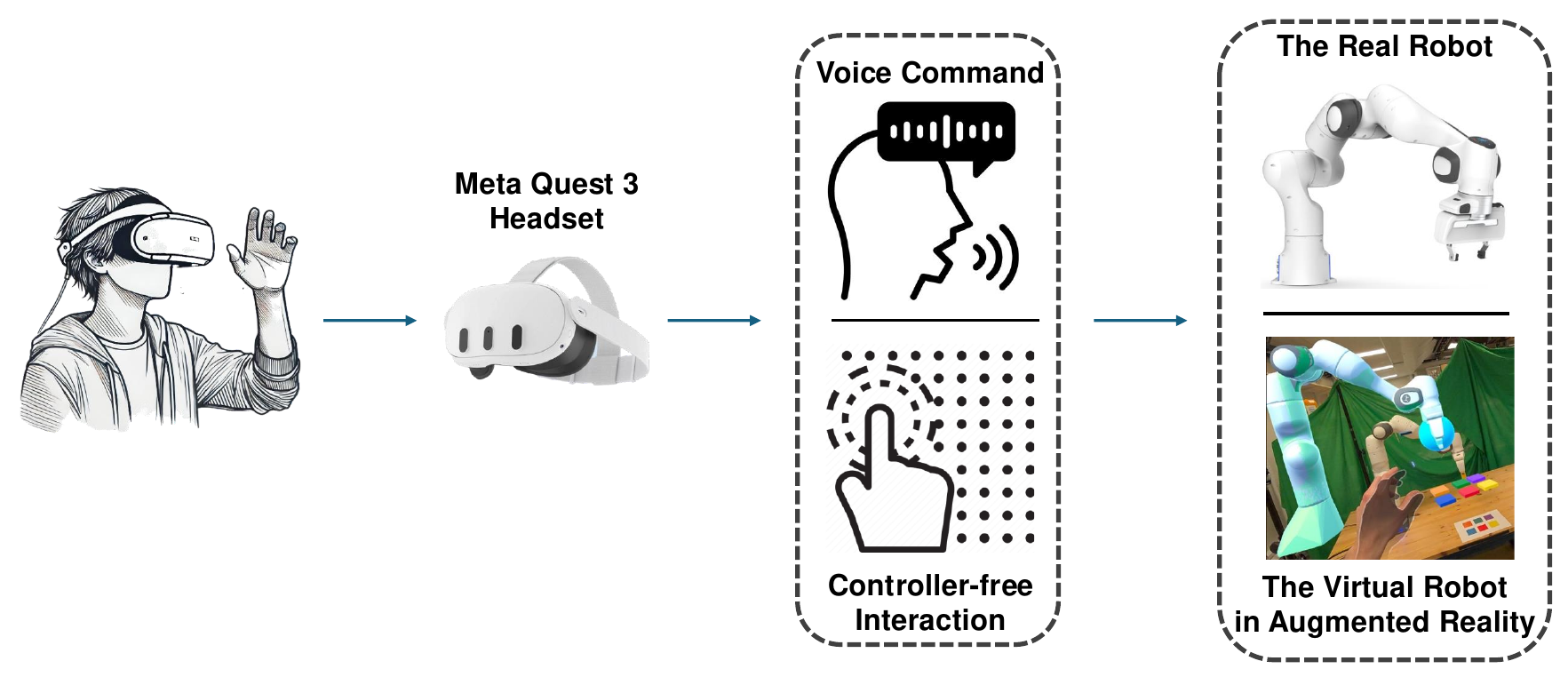}
\caption{An overview of our proposed LLM-driven, controller-free, voice-commanded AR robotic puppeteering system. The system operates within an AR environment, utilizing the Meta Quest 3 HMD. Users can interact seamlessly with the virtual robot, modeled identically after the real Franka robot arm using only hand gestures, while voice commands are integrated to enable intuitive control of the virtual robot, which in turn governs the real robot.}
\label{teaser}
\end{figure}

\begin{abstract}
The integration of robotics and augmented reality (AR) presents transformative opportunities for advancing human-robot interaction (HRI) by improving usability, intuitiveness, and accessibility. This work introduces a controller-free, LLM-driven voice-commanded AR puppeteering system, enabling users to teleoperate a robot by manipulating its virtual counterpart in real-time. By leveraging natural language processing (NLP) and AR technologies, our system—prototyped using Meta Quest 3—eliminates the need for physical controllers, enhancing ease of use while minimizing potential safety risks associated with direct robot operation. A preliminary user demonstration successfully validated the system’s functionality, demonstrating its potential for safer, more intuitive, and immersive robotic control.  

\end{abstract}
%

\section{Introduction}
Robotics has rapidly evolved over recent years, seamlessly integrating into various aspects of daily lives \cite{zhang2024will}, including service industries \cite{goel2020robotics,javaid2021substantial,dzedzickis2021advanced}, healthcare \cite{riek2017healthcare,sarker2021robotics,cresswell2018health}, and social scenarios \cite{breazeal2016social,tapus2007socially,baytas2019design}. This growing presence highlights the need for effective and intuitive human-robot interaction (HRI) \cite{zhang2024vision,ghiringhelli2014interactive,fang2012interactive} and collaboration \cite{green2008human,kyjanek2019implementation,de2020systematic}. A key enabler of this evolution is the integration of robotics with augmented reality (AR), a technology that superimposes virtual elements onto the real world to create immersive, context-aware interfaces \cite{zhang2023see,zhang2022site}. By bridging the gap between physical and digital realms, AR enhances usability, intuitiveness, and accessibility in human-computer interaction (HCI) \cite{zhang2023playing,zhang2023industrial}. Through its ability to present users with additional, contextually relevant information, AR provides transformative opportunities for advancing robotic teleoperation, situational awareness, and task performance \cite{sahin2024using,de2019intuitive,chan2022design}.

\begin{figure}
\centering
\includegraphics[width=.7\textwidth]{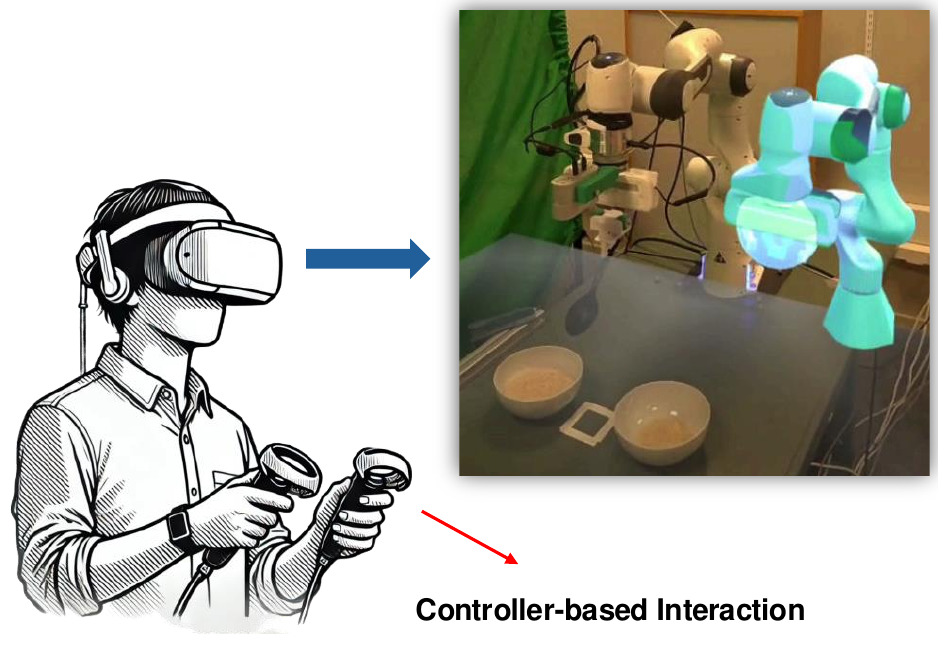}
\caption{The overview of the controller-based baseline AR robotic puppeteering system proposed by \cite{van2024puppeteer}.}
\label{van}
\end{figure}

Significant prior research has explored the application of AR in robotic teleoperation across diverse domains. These studies underscore AR's potential to improve user experience, enhance task execution, and bolster situational awareness – key components in achieving effective HRI \cite{arevalo2021assisting,walker2019robot,pan2021augmented,hedayati2018improving}. Among these contributions, the work of \cite{van2024puppeteer} stands out as particularly noteworthy. They introduced the first controller-based AR system for virtual robot teleoperation, enabling users to manipulate a virtual robot using physical controllers. Their innovative approach allowed the physical robot to replicate the trajectory of its virtual counterpart, laying a foundation for further advancements in AR-driven robotic teleoperation (Figure ~\ref{van}).

In parallel with the evolution of AR technologies, large language models (LLMs) have emerged as a transformative force across various fields, including robotics \cite{zhang2023large,zeng2023large}. Renowned for their superior natural language processing (NLP) capabilities, LLMs have been leveraged to enhance both robotic manipulation and interaction experiences \cite{gao2024physically,kim2024understanding}. Recently, in the realm of social robotics, LLMs facilitate intelligent conversational robots, enabling more natural and meaningful human-robot communication \cite{song2024vlm,lee2023developing}. Similarly, in robotic manipulation, these models have demonstrated their potential to improve task performance by providing advanced reasoning and decision-making capabilities \cite{li2024manipllm,chu2023accelerating,chen2024rlingua}. The integration of LLMs into robotic systems represents a promising avenue for redefining how humans interact with robots, paving the way for more intuitive, efficient, and accessible solutions.

Building on these advancements, this work introduces the LLM-driven AR Puppeteer system, a novel teleoperation framework that combines the power of LLMs and AR to enable voice-commanded, controller-free interaction with robots, as shown in Figure ~\ref{teaser}. Expanding upon the foundation established by \cite{van2024puppeteer}, our system eliminates the need for physical controllers, instead allowing users to intuitively manipulate a virtual robot through natural language instructions. These voice commands are processed in real-time by LLMs and seamlessly translated into actions executed by the physical robot. By replacing traditional controllers with voice-driven commands, the proposed system significantly enhances accessibility, making robotic teleoperation more inclusive for non-expert users and individuals with varying physical abilities. Our system is realized via Meta Quest 3, a new generation of optical see-through (OST) head-mounted display (HMD).

The fusion of cutting-edge NLP capabilities with immersive AR technologies offers an innovative approach to addressing challenges in HRI and HCI by introducing a controller-free, voice-driven interaction model supported by LLMs. This integration not only provides a groundbreaking framework for revolutionizing robotic teleoperation but also fosters a more natural and intuitive way for users to interact with robotic systems. By addressing barriers related to accessibility and ease of use, this human-centric approach emphasizes inclusivity and simplicity, broadening the reach of robotic applications across diverse domains \cite{zhang2024human,he2017survey,wallace2024imitation}. For instance, in industrial settings, workers can remotely command robots to perform complex tasks without specialized training \cite{norberto2005robot}. In healthcare, practitioners can rely on voice-guided teleoperation for precision in surgical robotics or patient care \cite{rogowski2022scenario,pulikottil2018voice}. Additionally, the system’s versatility can potentially be extended to fields like hazardous environment manipulation, educational robotics, and industrial automation, offering transformative possibilities for both expert and non-expert users \cite{zhang2024mind}.

The main contributions of this work are listed as follows:

\begin{itemize}
    \item Propose a novel and pioneering controller-free robotic teleoperation system leveraging AR and LLMs.
    \item Integrate LLM-driven voice command functionality within an OST HMD-based AR robotic teleoperation framework.
    \item Develop a direct and effective AR-driven robotic puppeteering system that enhances user accessibility and intuitiveness.
\end{itemize}

\section{Related Work}
Extensive research has explored the integration of extended reality (XR)—encompassing virtual reality (VR), augmented reality (AR), and mixed reality (MR)—into robotics across various domains. In this section, we review the current landscape of VR, AR, and MR applications in robotics, with a particular focus on AR for human-robot interaction (HRI).

\subsection{Virtual Reality in Robotics}
The use of XR technologies to facilitate robotics research has gained significant traction over the past decades. VR, in particular, has been widely adopted for simulating robots due to its ability to provide precise and controlled virtual environments \cite{sethi2009validation,lee2012validation,matsas2017design,whitney2019comparing}. Burghardt et al. \cite{burghardt2020programming} developed a VR system with digital twins to capture human movements in a virtual environment, enabling industrial robots to replicate complex human actions that are otherwise difficult to automate. Similarly, Togias et al. \cite{togias2021virtual} explored a VR-based teleoperation method for reprogramming industrial robots, enabling flexible and remote process design, which facilitates hybrid human-robot collaboration in dynamic production environments. 

However, VR’s integration with robotics has been particularly prominent in healthcare, especially for rehabilitation applications \cite{albani2007virtual,wade2011virtual,brutsch2011virtual,xia2024shaping}. Over three decades ago, Ojha \cite{ojha1994application} recognized VR’s potential in medicine and surgery, proposing its use for evaluating impaired hand performance, monitoring patient progress, and optimizing therapy strategies. Mirelman et al. \cite{mirelman2009effects} demonstrated that coupling robotic devices with VR can enhance post-stroke recovery by improving walking ability. Likewise, Frisoli et al. \cite{frisoli2007arm} developed a robotic exoskeleton integrated with VR for arm rehabilitation, showcasing its benefits in motor function recovery.

\subsection{Augmented Reality for Human-Robot Interaction}
The application of AR (often used interchangeably with MR) in HRI and robotic teleoperation \cite{solanes2020teleoperation,marin2002very,brizzi2017effects,zhao2017augmented} has become an established field. In recent years, AR has gained traction for robotic teleoperation due to its enhanced visualization capabilities and rich information presentation \cite{zhang2021supporting,nowak2021augmented,zhong2003designing}. For instance, Hedayati et al. \cite{hedayati2018improving} demonstrated how AR can improve robot teleoperation by providing intuitive visual feedback. Their user study on AR-based aerial robot interfaces highlighted significant usability and performance improvements over traditional methods that divide user attention between monitoring the robot and its camera feed. Additionally, Hashimoto et al. \cite{hashimoto2011touchme} introduced \textit{TouchMe}, a touch-based AR interface for intuitive robot teleoperation from a third-person perspective. Unlike traditional joystick controls, users could directly manipulate robot components by interacting with a live camera feed, significantly reducing the learning curve. In industrial settings, Michalos et al. \cite{michalos2016augmented} proposed an AR tool to support hybrid human-robot workplaces, offering real-time visualizations, instructional overlays, and production updates, thereby improving operator safety, acceptance, and workflow integration. For human-robot communication, Walker et al. \cite{walker2018communicating} explored AR-based motion intent cues to enhance interaction with non-anthropomorphic robots. Their study demonstrated that AR cues improve task efficiency and influence user perception of robots as teammates. 

Regarding AR-based robotic puppeteering, Van et al. \cite{van2024puppeteer} introduced the first optical see-through head-mounted display (HMD) AR system for robotic puppeteering, where users manually controlled a virtual robot through direct input. While this approach provided a foundation for AR-driven robot manipulation, it still required explicit manual control, limiting the fluidity and accessibility of the interaction. Our work builds upon and extends this concept by introducing a controller-free AR robotic puppeteering system, eliminating the need for physical controllers and enhancing usability. By integrating voice commands powered by LLMs, we enable a more natural and intuitive way for users to command and direct robot movements. This advancement not only improves interaction flexibility but also lowers the barrier for users with minimal experience, fostering a more seamless and immersive puppeteering experience in AR-driven robotic control.

\section{The System Introduction}

\subsection{Overview}
\begin{figure}[!t]
\centering
\includegraphics[width=\textwidth]{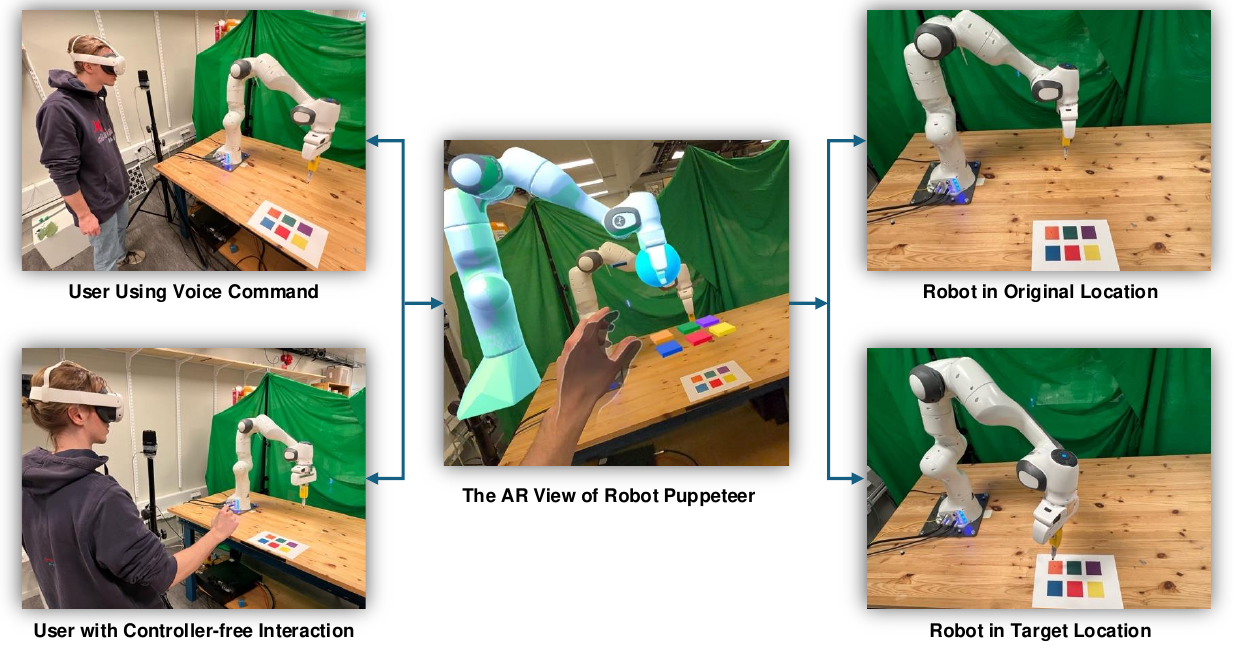}
\caption{A schematic overview of the proposed system with a user in action. The virtual Franka robot arm is rendered within the AR view (the middle one). Users can interact with the virtual robot through controller-free mid-air interaction by hands and voice commands signaled by a ZOOM H2N microphone (the left sub-figures), enabling precise manipulation of its movements (the right sub-figures). Simultaneously, the physical robot mirrors the virtual robot’s trajectory in real-time, completing the ‘puppeteer’ process seamlessly.}
\label{system}
\end{figure}

In terms of the equipment, this innovative system is built in the Meta Quest 3 HMD, which immerses users in a dynamic AR environment. The virtual robot arm, modeled after the real Franka robot arm \footnote{https://franka.de/}, serves as the primary interface for interaction and control. As shown in Figure ~\ref{system}, the physical robot operates in a practical setting, while following the identical trajectory of the virtual robot projected in the AR settings.

The proposed LLM-driven, voice-controlled, controller-free robotic puppeteering system follows a structured pipeline, as depicted in the figure. The system integrates user interaction, AR view, and real-time robotic puppeteer to create an intuitive and interactive teleoperation experience.

\begin{enumerate}
    \item \textbf{User Interaction:} This component highlights how users engage with the system through either voice commands or controller-free mid-air interactions. Voice commands, powered by LLMs, are processed to generate contextual movement instructions for the virtual robot. The controller-free hand interaction enables users to intuitively spawn the virtual robot at a desired location within the AR environment.
    
    \item \textbf{AR View:} Within the AR environment, the virtual robot arm acts as an intermediary, serving as a digital "puppet" modeled after the physical robot. Users define movements in AR, which are then transmitted to the real robot to be replicated. This mechanism ensures that all actions finalized in AR are accurately mirrored in the physical space, providing a seamless and interactive puppeteering process.
    
    \item \textbf{Real-time Robotic Puppeteering:} The physical robot executes tasks in real time, precisely replicating the movements defined by the virtual robot. In our current implementation, this process involves placing the robot from an initial position to a predefined target position. This final step completes the interaction loop, allowing users to visually confirm the execution of their voice commands within the AR view, while ensuring precise 'puppeteer' of the physical robot.
\end{enumerate}

\subsection{User Demonstration}
As illustrated in Figure ~\ref{system}, we conducted an initial demonstration with a small group of users from the authors’ local network. While this experiment successfully validated the core functionality of our proposed system, a more comprehensive pilot study and user evaluation remain to be designed and implemented.

The middle sub-figure in Figure ~\ref{system} showcases the AR interface projected in front of the user. Upon launching the dedicated application, the virtual robot was spawned alongside the user’s hand, following mid-air controller-free interactions. As depicted in the left-side sub-figures, the user issues voice commands via a connected microphone to control the virtual robot, eliminating the need for physical controllers. The entire process is achieved through a controller-free mechanism, enabling intuitive and hands-free interaction.

The robotic puppeteering is executed through the robot arm’s end-effector, which moves from its original position to a predefined target location, represented by a color-coded area, as shown in the right-side sub-figures. Once the virtual robot is spawned via hand interaction, the user issues a voice command -- “Blueberry, move to [color]” -- prompting the virtual robot’s end-effector to navigate toward the specified color. Simultaneously, the real robot mirrors this trajectory, precisely replicating the virtual robot’s movement in the real environment. Figure ~\ref{system} shows the example of a movement to the [orange] area.




\section{Technic Setup of the System}

This section provides a detailed description of the technological setup used in our proposed puppeteer system. The setup consists of three core modules: voice recognition, locally hosted LLMs, and the pipeline for handling transcriptions and commands.

\subsection{Voice Recognition}

For voice recognition and transcription, we utilized RealtimeSTT proposed by Kolja Beigel \footnote{https://github.com/KoljaB/RealtimeSTT}, a robust speech-to-text tool designed for real-time applications. The following configurations were applied:

\begin{itemize}
    \item \textbf{Wakeword:} We employ the wakeword "Blueberry"", which is one of the pre-trained options included in the RealtimeSTT framework. This wakeword triggers the voice recognition pipeline to start processing commands.
    \item \textbf{Speech Recognition Model:} The \texttt{tiny.en} model is used for lightweight and efficient transcription.
    \item \textbf{Wakeword Sensitivity:} The wake\_words\_sensitivity parameter is set to \texttt{0.9}, ensuring high sensitivity to the wakeword.
    \item \textbf{Microphone:} An external ZOOM H2N microphone is employed for audio input.
\end{itemize}

\subsection{Locally Hosted LLMs}

To process and validate commands, we run instances of locally hosted LLMs. This approach ensures low-latency interactions and data privacy. The setup includes:

\begin{itemize}
    \item \textbf{LLM Instances:} We use two instances of Llama 3.2 1B-Instruct Q6 models. These instances are hosted locally to maintain responsiveness and to avoid reliance on external servers.
    \item \textbf{Serving Framework:} The models are deployed using the \texttt{Llamafile} \footnote{https://builders.mozilla.org/project/llamafile/} framework, which provides an efficient and scalable mechanism for hosting multiple LLM instances on a local server.
\end{itemize}


\subsection{Pipeline for Handling Transcriptions}
Once the text is transcribed, it triggers multiple stages of validation and command handling, as detailed below:

\begin{enumerate}
    \item \textbf{Mode Identification:}
    The system first checks the current operational mode. If the mode is set to "puppeteer", the pipeline proceeds to look for specific "move-commands". Other modes may invoke different processing logic.

    \item \textbf{Command Validation:}
    Upon identifying potential commands, the system performs the following checks:
    \begin{enumerate}
        \item \textbf{Quick Validation:}
        The text is analyzed to determine if it contains the phrase "move to" followed by a predefined color (e.g., red, blue, orange, yellow, purple, green, or black (for tutorial only)). 
            \begin{itemize}
                \item If the command is valid ('yes'), the system returns a JSON object with the target color and proceeds to the next step.
                \item If the command is invalid ('no'), the system proceeds to the previous validation stage.
            \end{itemize}

        \item \textbf{LLM Validation:}
        Using a round-robin approach, the system gets the current prompt for further validation. The LLM performs the following assessments:
            \begin{itemize}
                \item If the command is identified as a valid "move-command" without any ambiguity, the system returns a JSON object with the target color and proceeds to the next step.
                \item If the command is clearly not a "move-command", the system discards it and return to the previous step.
                \item If the command is uncertain, the system marks the command for potential re-validation, which can be configured to custom rules.
            \end{itemize}
        \item \textbf{Command Execution:} The validated JSON object is sent to the OSC client via the UDP protocol.
\end{enumerate}
\end{enumerate}

\section{Discussion and Future Work}
In this section, we revisit our proposed system, reflect on its contributions, discuss its current limitations, and outline potential future work.

\subsection{Insights}
This work advances AR-based robotic teleoperation by introducing a controller-free, voice-driven interaction system powered by LLMs, significantly lowering barriers to robotic control. By addressing key HCI challenges -- such as accessibility, intuitiveness, and ease of use -- this system fosters natural, human-centric collaboration with robots. Unlike conventional robotic control methods or previous robotic puppeteering systems that rely on physical controllers or complex programming, our system enables users to engage with robots using context-aware voice commands and mid-air interactions. The AR interface improves usability by providing real-time visual feedback, allowing users to validate robotic movements instantly. Leveraging the Meta Quest 3 HMD, this system delivers an immersive robotic puppeteering experience, fostering greater confidence, precision, and engagement in robotic manipulation. This approach significantly reduces the learning curve for non-experts, making robotic teleoperation more intuitive, fluid, and accessible across various skill levels. By bridging advancements in NLP, AR, and robotics, this work lays the foundation for a more intuitive, flexible, and inclusive approach to robotic teleoperation.

A key strength of our system is its ability to decouple direct physical interaction from real-world robotic control, allowing users to manipulate a virtual robot while maintaining full situational awareness of the real robot. This approach minimizes risks associated with direct robotic operation, offering a safer and more intuitive method for teleoperation, particularly in high-stakes and industrial environments. By leveraging AR visualization and voice-driven commands, users can experiment with robot control in a low-risk virtual space, ensuring precision and confidence before executing real-world actions.

\subsection{Potential Practical Use}
The versatility of this system extends across multiple domains, offering practical solutions in industry, healthcare, hazardous environments, and education.

In industrial automation, it can streamline direct hands-on robotic control, enabling tasks such as assembly, maintenance, and quality inspection to be performed more efficiently and with less reliance on specialized training. For instance, in automotive manufacturing, workers could verbally instruct robotic arms to assemble components while monitoring their actions in real-time AR, reducing both training time and human error. Similarly, in warehouse logistics, workers could use gesture-based spawning and voice commands to remotely guide robotic pickers or conveyors, improving operational efficiency while reducing physical strain and workplace injuries. By minimizing direct physical engagement with heavy or hazardous machinery, this approach not only enhances productivity and accuracy but also significantly improves worker safety.

In healthcare, this technology has the potential to revolutionize assistive robotics, allowing individuals with limited mobility to interact with robotic systems through natural speech and AR visualization. This could be particularly impactful for stroke rehabilitation, where patients could guide a robotic exoskeleton simply by speaking commands, helping them regain motor function without requiring extensive physical effort. Similarly, in elderly care, voice-controlled robotic assistants could help with daily activities, such as fetching objects, adjusting room settings, or even assisting with physical therapy, all while allowing caregivers to monitor progress through an AR interface.

For hazardous environments, such as disaster response, nuclear facility management, and deep-sea exploration, voice-driven teleoperation eliminates the necessity for human presence in high-risk conditions. Emergency responders could, for example, remotely control robotic search-and-rescue units inside collapsed buildings while receiving live AR feedback on environmental hazards. In nuclear decommissioning, engineers could manipulate robotic arms to safely handle radioactive materials without exposure, using AR overlays to guide their precision. Similarly, in deep-sea exploration, researchers could issue high-level verbal commands to autonomous underwater vehicles, allowing them to conduct environmental monitoring, specimen collection, or infrastructure inspections without requiring expert-level piloting skills. By allowing operators to control robots from a safe distance, this system significantly enhances worker safety and operational effectiveness in some of the world’s most challenging environments.

Additionally, in education and research, our system provides an engaging and interactive platform for teaching robotic control principles without requiring coding or technical expertise. This can democratize access to robotics education, allowing students, educators, and researchers to experiment with robotic programming in an intuitive way. In practical classrooms, for example, students could manipulate virtual robots in AR, learning about kinematics, motion planning, and AI-driven control through hands-on interaction. Researchers in HRI could use this system to study intuitive command structures, user engagement, and learning curves, helping refine future robotic systems. By making robotics more accessible, this technology has the potential to foster innovation, inspire future generations, and accelerate advancements in HRI.

By enabling safe, efficient, and intuitive human-robot collaboration, this system bridges cutting-edge AR, NLP, and robotics technologies into a unified framework. Its broad applicability suggests that controller-free, voice-driven robotic control could play a pivotal role in shaping the future of automation, assistive technology, and intelligent robotics across industries.

\subsection{Limitations}
Despite the novel contributions, our system still has several limitations that must be addressed.  

First, voice recognition accuracy remains an area for improvement. During user demonstrations, we observed that in some cases, the system misinterpreted voice commands, especially under noisy conditions or when users had diverse accents or speech patterns. The effectiveness of the LLM-driven voice processing is dependent on further refinement to ensure more reliable and responsive recognition. These observed issues can also be addressed by using larger models for the voice recognition by adding more computational power.

Second, while our controller-free interaction allows users to spawn the virtual robot via mid-air interactions through hand tracking parameters, it does not yet support more advanced hand gestures for direct manipulation for users but only supports developers for hand gesture customization. The lack of robust gesture-based controls limits the system’s interactivity, which could otherwise enhance the overall user experience and precision of robotic control. 

Third, our robotic puppeteering implementation is currently limited to moving or navigating the robot arm through a single trajectory with basic motion planning. In real-world applications, robotic tasks often involve complex, multi-step motions that require more precise motion planning and more sophisticated voice command processing. Expanding the system to support dynamic, real-time trajectory planning is crucial for practical deployment in industrial and research settings.  

Lastly, although we conducted initial demonstrations with a small number of users, we have not yet performed a comprehensive user study with structured task design and systematic user evaluation. The lack of a formal user study limits our ability to quantitatively assess usability, efficiency, and user experience, which are essential for determining the system’s real-world practicality.

\subsection{Future Work}
To address these limitations and enhance the system’s capabilities, several aspects of future work are planned as follows.

First, we aim to improve voice recognition by integrating advanced speech models that are more robust to different accents, speech tones, and environmental noise. By using more powerful large models, we can make the voice command interpretation loop more inclusive and reliable.  

Second, we will expand gesture-based interactions beyond virtual robot spawning. By incorporating natural hand gestures for robotic control, users will be able to manipulate the virtual robot more precisely, providing a more fluid and interactive experience. This multimodal interaction will allow for a combination of voice and hand gestures, leading to more seamless robotic teleoperation.

Third, we plan to increase the complexity of robotic motion by incorporating more sophisticated voice commands that facilitate dynamic, context-aware motion planning, while also accounting for the robot’s degrees of freedom to enable more precise and adaptable control. This will allow robots to execute multi-step, adaptive movements, making them suitable for more complex tasks in industrial and assistive applications.  

Finally, we will conduct a comprehensive user study to rigorously evaluate the system’s usability, performance, and user experience. This study will involve a diverse group of participants across different expertise levels to assess the system’s effectiveness, intuitiveness, and practical applicability. Through user feedback, we will refine the interaction design, responsiveness, and overall workflow to ensure a more seamless and efficient robotic teleoperation system.  

By addressing these areas, we aim to push the boundaries of AR-driven robotic control, making it more practical, intelligent, safe, and accessible for a broad range of users across different industries.  

\section{Conclusion}
This work presents a controller-free, LLM-driven voice-commanded AR robotic teleoperation system, marking a significant step toward more intuitive, accessible, and immersive HRI. By integrating AR and NLP, our approach eliminates the constraints of traditional input devices, enabling fluid and natural communication between humans and robots. This advancement lowers the barrier to robotic control, making it more inclusive for users across different skill levels and application domains.  

Beyond enhancing accessibility and usability, this system establishes a versatile foundation for future developments in robotic teleoperation and interactive automation. Its potential extends to adaptive learning for task-specific customization, multimodal input integration, and collaborative robotics, where robots could intelligently respond to user commands in more dynamic and complex environments. Additionally, its ability to blend the physical and digital worlds through AR-driven visualization opens up new opportunities for safety, training, and real-time robotic supervision in industrial, medical, and other research settings.  


\section{Acknowledgment}
This work was supported by the Swedish Foundation for Strategic Research (SSF) grant FUS21-0067.

%
%
%
\bibliographystyle{splncs04}
\bibliography{Reference}
%




\end{document}